\documentstyle[12pt]{article} 

\begin{document}
\setcounter{page}{0}
\thispagestyle{empty}

\large
\centerline{\bf Shape Transition of Nanostructures created}
\vspace*{-0.2cm}
\centerline{\bf on Si(100) surfaces after MeV Implantation}
\normalsize
\vspace*{0.1cm}
\centerline {S. Dey$^{1}$, D. Paramanik$^{1}$, V. Ganesan$^{2}$ and S. Varma$^{1,*}$}
\vspace*{0.1cm}
\small
\centerline{\it ${^1}$Institute of Physics, Bhubaneswar - 751005, India.}
\vspace*{-0.3cm}
\centerline{\it ${^2}$IUC-DAEF, University Campus, Khandwa Road, Indore, India }
\normalsize
\vspace*{1.3cm}

\normalsize
\vspace*{1.3cm}
\begin{center}
{\bf Abstract}
\vspace*{-.4cm}
\end{center}

We have studied the modification in the Surface morphology of the 
Si(100) surfaces after 1.5~MeV Sb implantation. Scanning Probe
Microscopy has been utilized to investigate the ion implanted surfaces.
We observe the formation of nano-sized defect features on the Si 
surfaces for the fluences of 1$\times10^{13} ions/cm^2$ and higher. 
These nanostructures are elliptical in shape and inflate in size
for higher fluences. Furthermore, these nanostructures undergo a shape
transition from an elliptical shape to a circular-like at a
high fluence. We will also discuss the modification in surface roughness 
as a function of Sb fluence.

\vspace*{2.2cm}

\vspace*{0.2cm}
\noindent $^*$
 Corresponding author: shikha@iopb.res.in; 
Tel:91-674-2301058; FAX:91-674-2300142
 
\normalsize

\newpage
\section{Introduction}
\noindent

 In Si based semiconductor technology Sb is considered an important 
dopant for its role in the development of field effect transistors 
and infrared detectors \cite{temp}. Ion implantation is a useful 
technique for fabricating such devices as it produces buried layers 
with well defined interfaces, expanding possibility of designing 
novel structures. The increased density in VLSI circuits also makes 
the technological applications of the ion implantation , especially 
in MeV energy range, increasingly important. MeV implantation however 
can also produce severe modifications in the material depending on 
the nature and the energy of the impinging ion, and the implantation 
dose \cite{tam1}.
Extensive usage of ion implantation in device fabrication and the 
continued miniaturization of device structures has brought the 
issue of surface modifications, via ion implantations, to the 
forefront. However, the factors responsible for such modifications 
and the surface morphology after ion implantation, have received 
little attention \cite{car}.

 Atomic force microscope (AFM) is a very effective tool for 
examining surface modifications and surface structures. However,
there are very few studies in literature that have investigated 
the morphological changes of the ion implanted surfaces by AFM. 
Furthermore, most of these surface studies are performed after keV 
implantations \cite{cou,pia,wan}, or at low fluences for individual 
cascade studies \cite{wil}. Surface modifications after high energy,
100 MeV, ion irradiation have also been investigated \cite{sin}. 
However, the role of MeV ion implantation 
on the surface topography remains poorly understood. In the present 
study we have made detailed investigation on Si(100) surfaces after 
1.5~MeV Sb implantation. The technique of AFM has been applied to 
understand the modification in roughness and morphology of silicon 
surfaces upon ion implantation. We also investigate here the formation
 of nano-sized defects zones on Si(100) surfaces after MeV implantations.
 The results of shape transition in these nanostructures, from being
 elliptical at low fluences to becoming deformed circular at high fluences,
 will also be discussed here. Experimental procedure are mentioned in section~2
 and the results will be discussed in section~3. Conclusion are presented in section~4.
 
\section {Experimental}
\noindent     

A mirror polished (100)-oriented Si single crystal (p-type) 
wafer was used in the
present study. The samples were implanted at room temperature
with a scanned beam of 1.5 MeV Sb$^{2+}$ ions at various fluences
ranging from  1$\times 10^{11}$ to 5$\times10^{14} ions/cm^2$.
 The implantations were performed with
the samples oriented 7$^o$ off-normal to the incident beam to
avoid channeling effects. 

AFM Nanoscope E and Nanoscope III from Veeco were used to image
the implanted silicon sample surfaces. Images have been acquired
in contact and tapping modes. Images ranging from 0.2 to 10 $\mu{m}$ square 
were obtained.

\section {Results and Discussion}

Figure~1 shows 
the $1 \times 1\mu{m^2}$ and $200 \times 200 {nm^2}$ 
3D-images of the virgin silicon surface. It is observed that the
 virgin Si surface is smooth. 1.5 MeV implantation was carried out at 
 various fluences and several $1 \times 1\mu{m^2}$ and $200 \times 200 {nm^2}$ 
images were taken at all the fluences. The  $1 \times 1\mu{m^2}$ images 
were utilized for measuring the rms surface roughness of Si(100) surfaces
 after implantation.
The average roughness at each fluence is plotted 
in Figure~2. The rms roughness for a virgin Si(100) surface, 
measured to be 0.234~nm, is also marked. It can be clearly 
seen from Fig.~2 that the rms surface roughness exhibits 
three prominent behaviors as a function of fluence. For 
low fluences, up to $1\times10^{13} ions/cm^{2}$, the 
roughness is small and does not increase much compared to 
the virgin surface roughness. Beyond this fluence an enhanced 
surface roughness, increasing at a much steeper rate is observed. 
This trend continues up to the fluence of 
$1\times10^{14} ions/cm^{2}$ where a high roughness of 0.296~nm 
is measured. A saturation in surface roughness with a slight 
decrease in the roughness is observed beyond this fluence. 
The decrease in surface roughness, at 
$5\times10^{14} ions/cm^{2}$, seems reasonable in view of high 
level of amorphicity at this dose \cite{sdey2}, as beyond a certain 
high amorphicity, further higher levels of amorphization should 
tend to make the surface more homogeneous. A similar decrease 
in surface roughness with increasing fluence, beyond a critical 
fluence, has been observed for keV implantations of P and As 
in amorphous films \cite{edr}.

 Our earlier RBS/C and Raman scattering results 
\cite{sdey2,sdey3} show that Si lattice disorder also displays
3 similar behaviours as a function of ion fluence. Initially a low lattice 
damage, due to simple defects, is seen upto the fluence of 
$1\times10^{13}ions/cm^{2}$. The disorder becomes larger with the
onset of crystalline/amorphous (c/a) transition in Si-bulk 
at $1\times10^{13}ions/cm^{2}$. Finally the disorder 
saturates with the Si-lattice as well as Si-surface becoming 
amorphised at $5\times10^{15}ions/cm^{2}$.
The roughness on the Si
surface will be determined by several roughening and smoothening
process that undergo on an ion implanted surface. Nuclear Energy loss
effects are also crucial. In addition lattice disorder and the
associated stress will also be important in the evolution of the
ion implanted surface.

 High resolution $200\times200nm^{2}$ 
images of the Si-surfaces were acquired for all the fluences
and are shown in Figure~3 for two representative Sb fluences
of 1$\times10^{13}$ and 5$\times10^{14}ions/cm^2$. 
The images of Si surface acquired upto the  fluence of 
1$\times10^{12}ions/cm^2$ (not shown) are similar to the 
virgin surface (of Fig.~1b) and their surface roughness 
is also similar (Fig.~2). However, after a 
fluence of 1$\times10^{13}ions/cm^2$, several nanostructures
can be seen on the Si-surface (see Fig.~3a). Fig.~4a is same 
as Fig.~3a and shows the approximate outlines of some of the 
nanostructures. The nanostructures represent the damage due to 
ion implantation and are roughly of elliptical 
shape. For a quantitative analysis of 
these nanostructures, the two axial lengths 
were measured and the mean lengths of the minor and the major 
axes are found to be 11.6 and 23.0~nm respectively. The mean lengths 
of the two axes and the mean areas of the surface features 
are tabulated in Table~1 for various incident ion fluences. 

For a Sb fluence of 5$\times10^{13}ions/cm^2$, although the 
silicon surface is again found to be decorated with the 
elliptical nanostructures, the features have expanded
along both the axial directions (with the mean lengths 
of axes being 14.5 and 26.1~nm respectively). The average 
area of the nanostructures at this stage is calculated to be 
$297~nm^2$, which is about 41\% higher than that at 
1$\times10^{13}ions/cm^2$. For a fluence of 
1$\times10^{14}ions/cm^2$, the area of these  nanostructures
further inflates to $325\pm31~nm^2$. Although the length of 
the minor axis has not changed much at this fluence compared 
to that at 5$\times10^{13}ions/cm^2$, the major axis is elongated 
and has an average value of 31.6~nm.  Upto this stage the 
eccentricity of the elliptical structures, for all fluences 
is found to be $\sim$ 0.85$\pm$ 0.4. The eccentricity of the 
elliptical structures, at each fluence is listed in Table~ 1.  
Interestingly, after a fluence 5$\times10^{14}ions/cm^2$, the 
surface structures undergo a shape transition with the 
nanostructures having axial lengths of $30.1\pm4.4~nm$ and 
$30.7\pm2.4~nm$, respectively (see Fig.~3b). Fig.~4b is same 
as Fig.~3b and shows the approximate outlines of some of the 
nanostructures. The nanostructures have become much bigger in 
size and appear somewhat of circular shape. The nanostructures 
are not fully circular and have eccentricity $\sim$ 0.19$\pm$ 0.05 
(the eccentricity for circle=0). However, the eccentricity of these 
nanostructures is much reduced compared to those at lower fluences 
(where eccentricity $\sim$ 0.85$\pm$ 0.4). Hence, we refer to these 
nanostructures as {\it approximately circular}. An explosion in 
size  ($\sim$ 120\%) of these features compared to that at 
1$\times10^{14}ions/cm^2$ suggests a tremendous modification in 
surface morphology at this stage. Our results are in contrast to 
the keV implantation study of Sb in Si where, for doses lower 
than 1$\times10^{14}ions/cm^2$, no change in the surface topography 
was observed \cite{cou}.

The random arrival of ions on the surface constitute the stochastic
surface roughening. Surface diffusion, viscous flow, and surface sputtering
etc. contribute towards the smoothening of the surface \cite{eklund}.
The mechanism for the formation of surface damage is also postulated as a
results of cascade collision due to nuclear energy loss ($S_n$). In the 
present study also, $S_n$ seems to be the dominating factor in the creation of
the nanostructures after Sb implantation. In addition several factor like
c/a transition in Si lattice, the strain in the surface and in bulk, defect 
and disorder in the medium etc. may also responsible for the 
structure formation at the surface. For Sb implantation in Si we observe 
the formation of nanostructure at Si(100) surface only after the 
fluence of 1$\times10^{13}ions/cm^2$. These nanostructures inflate in size 
with increasing fluence. The size inflation may also be related to the increased 
disorder \cite{sdey2,sdey3} in the Si lattice. The shape transition of 
nanostructures, from being elliptical at lower fluence to deformed  circular at 
5$\times10^{14} ions/cm^2$ , may be caused by the increase in the density
of the electronic excitations. Our earlier studies \cite{sdey2,sdey3} 
show that the amorphization of Si-surface at this stage also leads to stress
relaxations on the ion implanted surface.

\section{Summary and conclusions}
\noindent

In the present study we have investigated  the modifications in 
the morphology of the Si(100) surfaces after 1.5~MeV Sb implantation. 
We observe the presence of nano-sized defect zones on the Si surfaces 
for the Sb fluences of 1$\times10^{13} ions/cm^2$ and higher.
These nanostructures are elliptical in shape and their size 
increase with fluence. We observe an abrupt
increase in size of nanostructures accompanied by a shape
transition after the fluence of  5$\times10^{14}ions/cm^2$.  
The nanostructures become approximately circular at this stage. 
We have also investigated the modifications in the surface
roughness of the ion implanted Si surfaces and find that
surface roughness demonstrates 3 different stages as a function 
of fluence. 

\section{Acknowledgments}
\noindent

This work is partly supported by ONR grant no. N00014-97-1-0991.
We would like to thank A.M. Srivastava for very useful comments 
and suggestions. We would also like to thank Puhup Manas
for his help with the figures.

 \newpage
\noindent{\bf \Large {Figures}}

\vskip 0.3 in
\noindent Fig. 1:  3-D AFM images of the virgin Si(100) surfaces.
Image areas are (a) $1 \times 1\mu{m^2}$ 
and (b) $200 \times 200~{nm^2}$ . 

\vskip 0.3 in
\noindent Fig. 2: The rms roughness of the Sb implanted Si(100)
surfaces, measured using AFM, is plotted as a function of Sb ion
fluence. Symbol sizes denote the error in the measurement. Data 
for the virgin sample is also shown. 

\vskip 0.3 in
\noindent Fig. 3: Surface structures on silicon surface:
AFM images of silicon surfaces after implantation
with 1.5~MeV Sb ions at the fluences of (a) $1\times 10^{13} ions/cm^2$,
and (b) $5\times 10^{14} ions/cm^2$.  Area of each image is
$200 \times 200~nm{^2}$. 

\vskip 0.3 in
\noindent Fig. 4: (a)Same as Fig.~3a with approximate
outlines of some of the nanostructures drawn and (b) same as Fig.~3b
with approximate outlines of some of the nanostructures drawn.

\vskip 0.3 in

\noindent{\bf \Large {Table}}
\vskip 0.3 in
\noindent Table 1: Lengths of minor axis, major axis, 
area and the eccentricity of the nanostructures 
as a function of Sb ion fluence.

\newpage

\thispagestyle{empty}
\vspace*{0.8cm}
\noindent{\bf \Large {Table1}}

\vspace*{1.8cm}

\begin{tabular}{|c|c|c|c|c|}\hline\hline
\rm Fluence        &\rm Minor Axis  &\rm Major Axis   &\rm  Area            &\rm  Eccentricity \\
\rm ($ions/cm^2$)    &\rm (nm)            &\rm (nm)   &\rm $(nm^2)$         &\rm       \\
\hline  
\rm $1\times10^{13}$ &\rm 11.6$\pm 2.2$ &\rm 23.0$\pm 3.7$ &\rm 210$\pm23$  &\rm 0.86$\pm0.04$\\
\rm $5\times10^{13}$ &\rm 14.5$\pm 1.8$ &\rm 26.1$\pm 2.9$ &\rm 297$\pm22$  &\rm 0.83$\pm0.04$\\
\rm $1\times10^{14}$ &\rm 13.1$\pm 2.3$ &\rm 31.6$\pm 4.2$ &\rm 325$\pm31$  &\rm 0.87$\pm0.05$\\
\rm $5\times10^{14}$ &\rm 30.1$\pm 4.4$ &\rm 30.7$\pm 2.4$ &\rm 726$\pm52$  &\rm 0.19$\pm0.05$\\
\hline\hline
\end{tabular}


\newpage
\begin{thebibliography}{99}

\bibitem{temp} G. Tempel, N. Schwarz, F. Muller, F. Koch, H.P. Zeindl,
               and I. Eisele, Thin Solid Films {\bf 174} (1990) 171.
\bibitem{tam1} M.Tamura and T. Suzuki, Nucl. Instr. and Meth. 
               {\bf B39} (1989) 318.
\bibitem{car} G. Carter, M.J. Nobes, I.V. Katardjiev and J.L. Whitton
              Defect and Diffusion Forum, {\bf vol.57/58} (1988) 97-126.
              Ion Implantation 1988, ed. F.H. Wohlbier (Trans. Techn. 
              Publ. Ltd).
\bibitem{cou} D. Courboin, A. Grouillet, E. Andre, Surf. Sci. {\bf 342}
              (1995) L1111.
\bibitem{pia}  A. Piatkowska, G. Gawlik and J. Jagielski, Appl. Surf. 
               Sci, {\bf 141} (1999) 333. 
\bibitem{wan} J.B. Wang, A.Datta, and Y.L. Wang, Appl. Surf. Sci.
              {\bf 135} (1998) 129.
\bibitem{wil} I.H. Wilson, N.J. Zheng, U. Knipping, and I.S.T. Tong,
              Phys. Rev. B. {\bf 38} (1988) 8444.
\bibitem{sin} J.P. Singh, R. Singh, N.C. Mishra, D. Kanjilal, V. Ganesan, 
              J. Appl. Phys. {\bf 90} (2001) 5968.
\bibitem{sdey2} S. Dey, C. Roy, A. Pradhan and  S. Varma, J. Appl. Phys. 
                {\bf 87} (2000) 1110.
\bibitem{edr} R. Edrei, E.N. Shauly, A. Hoffman, J. Vac. Sci. Tech. A
                {\bf 20} (2002) 344. 
\bibitem{sdey3} S. Dey, A. Pradhan and  S. Varma, J. Vac. Sci. Tech. B 
                {\bf 18} (2000) 2457. 

\bibitem{eklund} E.A. Eklund, E.J. Snyder, and R.S. Williams. Surf. Sci.
              {\bf 285} (1993)1571.

\end{thebibliography}
\end{document}